# Assessing Modeling Fidelity for Long-Term Battery Energy Storage Planning: Operation, Degradation, and Temporal Representation

Hassan Zahid Butt and Xingpeng Li

*University of Houston, 4222 Martin Luther King Blvd, Houston, Texas 77204, USA*

## Abstract

Long-term battery energy storage system (BESS) planning often relies on simplified degradation, operational, and temporal representations to maintain computational tractability, yet their effects on lifecycle conclusions are not well understood. This paper assesses the modeling fidelity needed for long-term lifecycle evaluation of BESS designs used in planning studies. A 20-year grid-connected microgrid is sized using a degradation-naive planning model, after which the installed portfolio is fixed and evaluated through sequential lifecycle validation. The reference representation combines nonlinear calendar and cycle aging, C-rate-dependent efficiencies, state-of-health-dependent performance, self-discharge, battery replacement, and full 8,760-h chronology. Battery-model hierarchies, targeted ablations, linear degradation surrogates, health-update intervals, temporal reductions, and combined simplifications are compared using lifecycle cost, replacement timing, state of health, and energy adequacy. The reference case produces replacements in years 9 and 18, a $111.25 million lifecycle net present cost, and 24.01 MWh of cumulative energy not served. Omitting calendar aging eliminates both replacements and understates lifecycle cost by 35.1%, whereas a separately calibrated linear surrogate model reproduces both replacement years and limits the cost deviation to 0.2%, although energy not served remains 31.8% below the reference. A peak-informed calibrated 12-day representation preserves replacement timing but reports zero energy not served, while replacing the preserved peak day with the maximum daily-energy-deficit day substantially overstates energy not served because representative-day closure alters the battery state surrounding the critical event. The results show that modeling fidelity is metric-dependent and should be selected according to the lifecycle outcome to be preserved.

## Nomenclature

Sets and indices:

| | |
|---|---|
| $Y, y$ | Set and index of planning years |
| $\mathcal{T}, t$ | Set and index of hourly operating periods |
| $B, b$ | Set and index of battery health update intervals |
| $\mathcal{D}, d$ | Set and index of calendar days |
| $\mathcal{T}_b$ | Set of hourly periods within health interval $b$ |
| $i, j$ | Indices for C-rate breakpoint, and rainflow cycle index |
| $K^{rep}, k$ | Set and index of battery replacement events |
| $\mathcal{R}, \rho$ | Set and index of representative operating days |
| $\Omega_j^{act}$ | Active charging/discharging periods of a cycle $j$ |
| <u>Parameters:</u> | |
| $P_{y,t}^{L}$ | Electrical demand (MW) |
| $\phi_t^{PV}$ | Year-1 PV availability factor (p.u.) |
| $g_L$ | Annual load growth rate |
| $\delta^{PV}$ | Annual PV degradation rate |
| $P^{PCC}$ | Grid interconnection capacity (MW) |
| $c^{PV}, c^{B}$ | PV and BESS capital cost factors (\$/MW and \$/MWh) |
| $c_{y,t}^{imp}, c_{y,t}^{exp}$ | Grid import and export prices (\$/MWh) |
| $c_{\mathrm{FOM}}^{PV}$ | Annual PV fixed O&M cost coefficient (\$/MW-year) |
| $r$ | Real discount rate |
| $\gamma$ | Reliability enforcement coefficient (\$/MWh) |
| $\epsilon$ | Numerical threshold for identifying unserved load (MW) |
| $\Delta t$ | Operating interval duration (h) |
| $\Delta t_s$ | Calendar-aging interval duration (s) |
| $h^B$ | Rated BESS duration (h) |
| $P^{B,max}$ | Rated BESS charging/discharging power (MW) |
| $SOH^{EOL}$ | End-of-life (EOL) BESS replacement threshold (p.u.) |
| $\sigma_0$ | BOL hourly self-discharge rate |
| $\eta_0^{ch}, \eta_0^{dch}$ | BOL charging and discharging efficiencies |
| $T, T_{ref}$ | Cell and reference temperatures (K) |
| $R$ | Universal gas constant (kJ/(mol·K)) |
| $E_a$ | Calendar-aging activation energy |
| $k_{ref}, c_Q, d_Q$ | Calendar-aging coefficients |
| $z$ | Degradation exposure exponent |
| $k_T, k_{SOC}$ | Calendar-aging temperature and SOC stress coefficients |
| $k_C, k_{DoC}$ | Cycle-aging C-rate and depth-of-cycle stress coefficients |
| $k_T^{cyc}$ | Cycle-aging temperature stress coefficient |
| $n_j$ | Rainflow cycle count |
| $DoC_j$ | Depth of cycle $j$ |
| $C_i$ | C-rate at breakpoint $i$ |
| $s_\eta$ | BOL efficiency normalization factor |
| $m_\eta$ | SOH-dependent efficiency multiplier |
| $s_b^m, s_0^m$ | Monthly and BOL monthly self-discharge fractions |
| $\sigma_b$ | Hourly self-discharge rate during health interval $b$ |
| $w_\rho$ | Annual weight of representative day $\rho$ |
| <u>Planning, operating and battery state variables:</u> | |
| $S^{PV}, S^{B}$ | Installed PV (MW) and BESS capacity (MWh) |
| $P^{imp}, P^{exp}$ | Grid import and export power (MW) |
| $P^{PV}, P^{PVcurt}$ | Utilized and curtailed PV power (MW) |
| $P^{ch}, P^{dch}$ | BESS charging and discharging power (MW) |
| $P^{shed}$ | Unserved load (MW) |
| $E^B$ | Stored BESS energy (MWh) |
| $SOC_t$ | BESS state of charge at operating period $t$ (p.u.) |
| $SOH_b$ | BESS state of health during interval $b$ (p.u.) |
| $E_b^{avail}$ | Available BESS energy capacity during interval $b$ (MWh) |
| $C_{b,t}$ | Instantaneous BESS C-rate |
| $C_j^{eff}$ | Effective C-rate of rainflow cycle $j$ |
| $P_{b,i}$ | Power corresponding to breakpoint $i$ during interval $b$ |
| $G_{b,t}, W_{b,t}$ | Effective stored-energy gain and withdrawal rates (MW) |
| $G_{b,i}, W_{b,i}$ | Stored-energy gain and withdrawal rates at breakpoint $i$ (MW) |
| $\lambda_{b,t,i}^{ch/dch}$ | SOS2 interpolation weights for charging and discharging |
| Degradation and lifecycle quantities: | |
| $Q_b^{cal}, Q_b^{cyc}, Q_b^{tot}$ | Calendar, cycle, and total capacity loss (%) |
| $t_t^*$ | Equivalent calendar-aging exposure under stress condition |
| $FEC, \Delta FEC_j$ | Full equivalent cycles (FEC), and FEC contribution of cycle $j$ |
| $FEC_j^*$ | Equivalent cycling exposure under the current cycle stress |
| $Q_{loss}^{lin}$ | Capacity loss predicted by linear degradation surrogate (%) |
| $\mathrm{FEC}^{throughput}$ | Throughput-based equivalent full cycles |
| $\alpha_{\mathrm{cal}}$ | Linear calendar-aging coefficient (%/h) |
| $\alpha_{\mathrm{cyc}}$ | Linear cycling-throughput coefficient (%/FEC) |
| $T_{\mathrm{age}}$ | Elapsed battery age since commissioning (h) |
| $C_y^{op}$ | Annual operating cost in year $y$, including net grid transactions and PV fixed O&M ($) |
| $C_k^{rep}$ | Cost of battery replacement event $k$ ($) |
| $\tau_k$ | Time of replacement event $k$, measured from project start year |
| $\mathcal{H}_d$ | Set of hourly periods belonging to calendar day $d$ |
| $D_d$ | Daily energy deficit above the PCC limit on day $d$ (MWh) |
| $ENS, LOLH$ | Energy not served (MWh), and loss of load hours (h) |
| $N^{SF}, P_{max}^{SF}$ | Number of shortfall days (days), and peak hourly shortfall (MW) |
| $NPC$ | Lifecycle net present cost ($) |

# 1. Introduction

Battery energy storage systems (BESSs) are increasingly used in microgrids and distributed energy systems for energy shifting, renewable integration, peak-demand management, and backup supply [1]. Their long-term planning value depends on installed capacity and evolving battery performance. Capacity loss reduces dispatchable energy, while efficiency and self-discharge affect the energy that can be stored and returned. Over project horizons that may exceed a battery service life, these effects influence both lifecycle cost and energy adequacy.

Representing this behavior in long-term planning models is challenging because battery condition depends on the operating trajectory produced by the optimization itself. Calendar aging varies with time, temperature, and state of charge (SOC), while cycle aging depends on cycle depth, operating rate, temperature, and SOC [2]–[5]. Charging and discharging efficiencies can also vary with operating rate and battery condition [6]. These dependencies create feedback among hourly dispatch, degradation, available energy capacity, operating losses, and replacement timing. Fig. 1 provides a general taxonomy of degradation and operational dependencies relevant to long-term BESS representation.

The level of detail that can be retained is constrained by computational tractability. Long-term BESS planning studies have incorporated degradation through simplified aging relationships, rainflow-based cycle models, sequential health updates, and replacement formulations [7]–[11]. Temporal reduction introduces a related tradeoff. Full hourly chronology preserves the sequence of demand, renewable generation, prices, and storage states, whereas representative periods substantially reduce problem size [12], [13]. For storage, however, temporal aggregation can disrupt state continuity or omit operating conditions that determine energy adequacy [12], [14]. A reduced model may therefore reproduce annual energy or lifecycle cost while failing to preserve short-duration adequacy events.

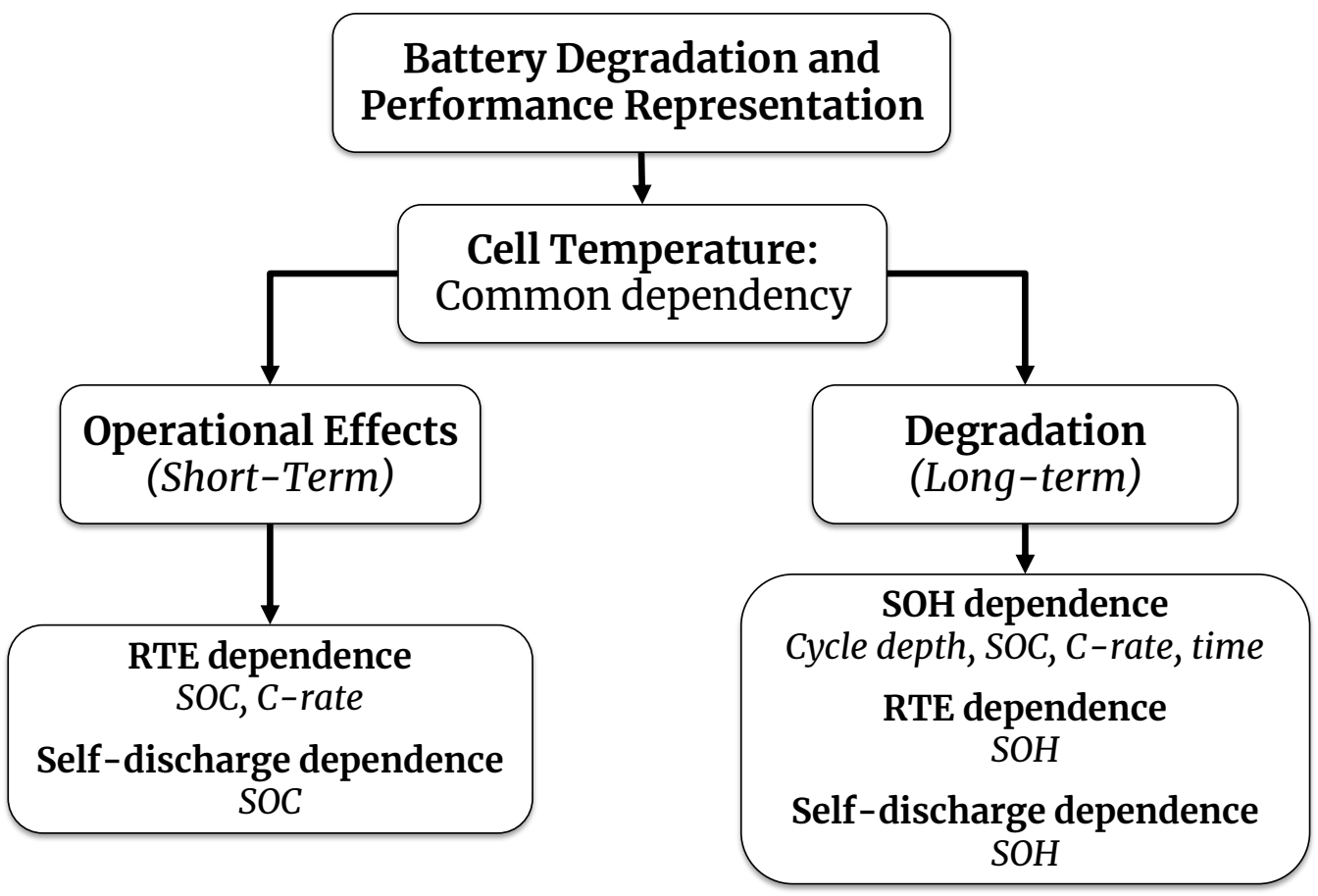

Fig. 1. Battery degradation and operational performance dependencies. RTE: round-trip efficiency; SOH: state of health.

The appropriate modeling fidelity is therefore not necessarily the same for every planning outcome. Additional battery detail may provide limited value under mild operating conditions, while apparently small simplifications can become important when the system operates near an adequacy boundary. Fidelity should consequently be assessed against the particular economic, degradation, or adequacy metric that a planning study intends to preserve. This paper investigates this issue through a controlled 20-year grid-connected microgrid study. A degradation-naive planning model first determines photovoltaic (PV) and BESS capacities, after which the installed portfolio is fixed and evaluated sequentially under alternative degradation, operational, health-update, and temporal representations. The study examines which battery mechanisms materially affect lifecycle cost, state of health (SOH), replacement timing, and energy adequacy; how health-update frequency influences lifecycle outcomes; and how temporal reduction affects the preservation of adequacy-critical operating conditions. Rather than seeking the most detailed possible battery model, the objective is to determine which modeling details are consequential for the lifecycle quantity being evaluated.

## 2. Literature Review

### *2.1 Battery Degradation and Operational Representation*

Battery degradation has been represented in long-term planning at different levels of physical and computational detail. Linear and throughput-based formulations remain attractive because they can be embedded in mixed-integer linear programs, while more detailed approaches include depth-dependent degradation, rainflow cycle counting, sequential health updates, and replacement modeling [8]–[11]. Their reported impact on storage sizing, replacement timing, and lifecycle economics depends strongly on operating duty and degradation representation.

More detailed models retain stress factors such as calendar exposure, cycle depth, SOC, temperature, and operating rate. Calendar and cycle degradation have been represented jointly using semi-empirical relationships [15], while current derating and stress-factor approaches capture SOC-, temperature-, and rate-dependent effects [16]. Rainflow-based and piecewise-linear formulations provide a compromise between cycle-level fidelity and optimization tractability [10], [17], while data-driven degradation models have also been incorporated through decomposition or external solution procedures [7]. Reviews similarly show continued use of empirical and semi-empirical models because they retain important aging mechanisms at lower computational cost than electrochemical models [18], [19].

Long-term planning models often simplify battery performance more aggressively than capacity degradation. Fixed charging and discharging efficiencies preserve linearity, although experimental evidence shows that conversion losses vary with operating rate and can be represented using piecewise-linear one-way efficiency relationships [6]. Calendar aging may dominate under low-C-rate operation [16], whereas rate-dependent effects become more important at higher operating power. Battery aging can also affect efficiency and self-discharge, although transferable quantitative relationships are less established than for capacity fade [20]. These studies show that different battery mechanisms affect different lifecycle outputs. Capacity fade directly affects available energy and replacement, whereas efficiency and self-discharge primarily affect the operating energy balance. A model that reproduces battery lifetime or lifecycle cost may therefore not preserve energy adequacy, motivating controlled comparison of individual mechanisms using a common installed portfolio.

### *2.2 Temporal Representation in Long-Term Planning*

Temporal resolution introduces a separate tradeoff because BESS operation depends on chronological state evolution. Full hourly simulation preserves the sequence of demand, renewable generation, prices, and stored energy, but becomes computationally expensive over multi-year horizons. Representative periods, clustering, and time-series aggregation are therefore widely used to reduce problem size [12], [13].

Storage is particularly sensitive to temporal aggregation because energy states link operating periods. Independent representative periods can misrepresent storage behavior when inter-period energy transfer is important, motivating chronology-preserving formulations and representative-day methods that retain transitions, extreme conditions, or additional temporal information [12], [14], [21], [22]. Preserving annual energy totals alone is therefore insufficient when storage value depends on temporal sequencing.

Approximation quality is commonly assessed using system cost, installed capacity, energy production, or storage utilization, but these metrics may not reveal whether infrequent adequacy-critical conditions are retained. For a storage-constrained system, the highest instantaneous net-load hour may not occur on the day with the largest sustained energy deficit. Reduced chronology can therefore reproduce annual quantities and degradation behavior while omitting the sequence responsible for energy not served (ENS). Temporal fidelity should consequently be evaluated using adequacy metrics as well as economic and degradation outcomes.

### *2.3 Research Gaps and Contributions*

Existing studies establish that battery degradation, operational performance, and temporal representation can each affect long-term storage planning, but these elements are commonly evaluated using different system designs, degradation formulations, health-update structures, or temporal models. Their effects are therefore difficult to separate directly. Fidelity is also often judged using cost, installed capacity, or battery lifetime, even though lifecycle cost, SOH, replacement timing, ENS, loss of load hours, shortfall frequency, and peak shortfall may respond differently to the same approximation.

A further gap concerns interactions among simplifications: degradation studies often adopt a fixed temporal representation, while temporal-aggregation studies commonly use simplified battery physics. A controlled assessment using the same installed portfolio is therefore needed to distinguish the effects of degradation, operational, health-update, and temporal fidelity across different lifecycle outcomes. Accordingly, this paper makes four contributions:

1. A controlled long-term lifecycle fidelity assessment compares degradation, operational, health-update, and temporal representations using a common fixed PV-BESS portfolio and reference case, separating representation effects from changes in optimized system design.
2. A literature-derived composite battery representation combines nonlinear calendar aging, rainflow-based cycle aging, cycle-specific C-rate effects, C-rate-dependent efficiencies, SOH-dependent performance, self-discharge, and replacement. A progressive hierarchy and targeted ablations separate cumulative from mechanism-specific effects.
3. Health-update intervals and alternative temporal representations are evaluated using economic, degradation, replacement, and energy-adequacy metrics to determine whether agreement in lifecycle cost or SOH also preserves event-sensitive adequacy outcomes.
4. Two linear degradation surrogates and a combined simplification case distinguish degradation-mechanism omission from linearization, test the transferability of first-life calibration, and quantify interactions among individually modest approximations.

# 3. Methodology

## *3.1 Planning and Lifecycle Assessment Framework*

The methodology separates long-term capacity planning from degradation-aware lifecycle assessment through the four-stage procedure in Fig. 2. A degradation-naive planning model first determines PV and BESS capacities over the 20-year horizon while representing annual load growth and PV degradation. The BESS uses fixed beginning-of-life (BOL) efficiencies and self-discharge, with no capacity fade or replacement.

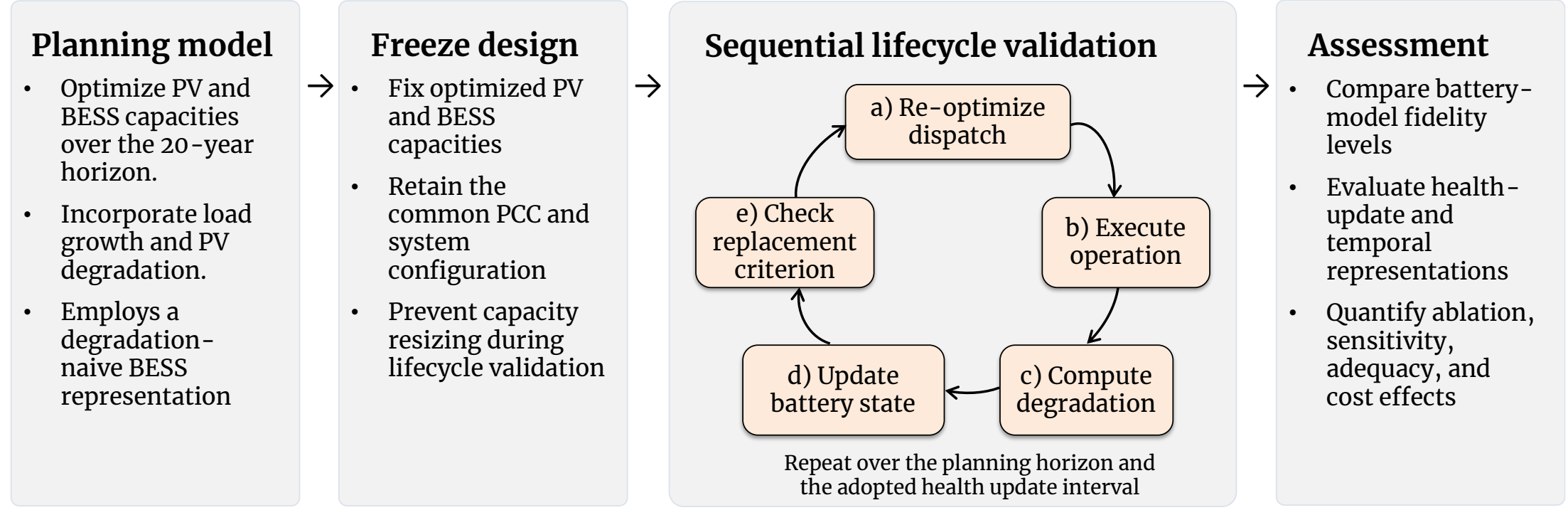

Fig. 2. Long-term BESS fidelity assessment framework.

The optimized PV and BESS capacities are then fixed for all lifecycle experiments. Operation is evaluated sequentially over battery health-update intervals; executed dispatch determines calendar and cycle degradation, after which state of health (SOH), available energy, efficiency, and self-discharge are updated before the next interval. When the SOH replacement threshold is reached, a replacement is recorded and the degradation state is reset.

Lifecycle trajectories are compared under alternative battery and temporal representations using economic, degradation, replacement, and adequacy metrics. Fixing the installed portfolio ensures that differences among cases reflect modeling fidelity rather than changes in optimized system design.

### *3.2 Long-Term Planning Formulation*

The first stage determines installed PV and BESS capacities using a multi-year grid-connected microgrid planning model. The planning horizon is indexed by $y$, with hourly operating periods indexed by $t$. The objective minimizes PV and BESS investment and discounted operating cost, including net grid transactions and PV fixed O&M:

$$\min \; c^{PV}S^{PV} + c^{B}S^{B} + \sum_{y\in Y}\frac{1}{(1+r)^{y-1}}\left[\sum_{t\in\mathcal{T}}\left(c_{y,t}^{imp}P_{y,t}^{imp} - c_{y,t}^{exp}P_{y,t}^{exp} + \gamma P_{y,t}^{shed}\right)\Delta t + c_{\mathrm{FOM}}^{PV}S^{PV}\right] \tag{1}$$

where $r$ is the discount rate and $\gamma$ is a large reliability-enforcement coefficient used to discourage avoidable load shedding while preserving feasibility. The coefficient is introduced for numerical enforcement and is not interpreted as a value of lost load.

Hourly supply and demand are balanced according to

$$P_{y,t}^{imp} + P_{y,t}^{dch} + P_{y,t}^{PV} + P_{y,t}^{shed} = P_{y,t}^{L} + P_{y,t}^{ch} + P_{y,t}^{exp} \tag{2}$$

Grid import and export are each bounded by the interconnection capacity $P^{PCC}$, while mutually exclusive import/export and charge/discharge operation are enforced through binary operating variables. Annual load growth is represented as

$$P_{y,t}^{L} = P_{1,t}^{L}(1+g_L)^{y-1} \tag{3}$$

where $g_L$ is the annual load growth rate. Available PV generation decreases annually as

$$P_{y,t}^{PV} + P_{y,t}^{PVcurt} = S^{PV}\phi_t^{PV}(1-\delta^{PV})^{y-1} \tag{4}$$

where $S^{PV}$ is the installed PV capacity, $\phi_t^{PV}$ is the year-1 PV availability profile, and $\delta^{PV}$ is the annual PV degradation rate. During planning, the BESS is represented using its rated energy capacity $S^B$, with $0 \le E_{y,t}^{B} \le S^B$. Stored energy evolves using fixed BOL efficiency and self-discharge:

$$E_{y,t+1}^{B} = (1-\sigma_0)E_{y,t}^{B} + \eta_0^{ch}P_{y,t}^{ch}\Delta t - \frac{P_{y,t}^{dch}\Delta t}{\eta_0^{dch}} \tag{5}$$

where $\eta_0^{ch}$ and $\eta_0^{dch}$ are the fixed beginning-of-life charging and discharging efficiencies used in the planning model. The adopted system-level BOL RTE of 0.85 is split symmetrically between charging and discharging, while $\sigma_0$ denotes the fixed hourly self-discharge rate. Charging and discharging power are bounded by $P^{B,\max}$, which follows the adopted fixed-duration architecture:

$$P^{B,\max} = \frac{S^B}{h^B} \tag{6}$$

where $h^B$ is the rated BESS duration. The resulting $S^{PV}$ and $S^B$ are fixed during all lifecycle fidelity experiments. The planning model is initialized at 50% SOC and maintains chronological BESS energy continuity across all hourly periods. A 50% terminal SOC condition is imposed at the end of each modeled year to prevent inter-year energy shifting; no daily or intra-year cyclic SOC closure is imposed.

### 3.3 Battery Degradation and Replacement

Battery degradation is evaluated from the executed BESS trajectory using complementary calendar- and cycle-aging relationships derived from the same commercial LFP/graphite cell family [4], [5]. Degradation is calculated after each executed operating interval rather than embedded directly in the dispatch optimization, retaining stress-dependent aging without introducing the semi-empirical equations into the mixed-integer formulation.

#### 3.3.1 Calendar Aging

Calendar-induced capacity loss follows the semi-empirical relationship in [4]:

$$Q_b^{cal} = k_T(T)\, k_{SOC}(SOC)\, t^z, \qquad z = 0.5 \tag{7}$$

where $Q_b^{cal}$ is the accumulated calendar capacity loss in percent, $T$ is cell temperature, $SOC$ is the normalized state of charge, and $t$ is the corresponding calendar exposure in seconds. The temperature and SOC dependence is represented by

$$k_T(T) = k_{ref} \exp\left[-\frac{E_a}{R}\left(\frac{1}{T} - \frac{1}{T_{ref}}\right)\right] \tag{8}$$

$$k_{SOC}(SOC) = c_Q(SOC - 0.5)^3 + d_Q \tag{9}$$

The adopted coefficients are represented in Table 1. Battery temperature is fixed at $25°\mathrm{C}$ in this study, so the temperature coefficient remains at its reference value while SOC varies according to the executed hourly trajectory.

Table 1 Degradation model parameters from [4]

| ***Parameter*** | ***Value*** | ***Description*** |
|---|---|---|
| $T_{ref}$ | 298.15 K | Reference temperature |
| $k_{ref}$ | 0.0012571 %/$\sqrt{s}$ | Calendar aging coefficient |
| $E_a$ | 17.126 kJ/mol | Activation energy |
| $c_Q$ | 2.8575 | SOC coefficient |
| $d_Q$ | 0.60225 | SOC offset |

Because SOC-dependent stress changes during operation, applying the current stress coefficient to total elapsed age would incorrectly reassign prior degradation to the present SOC condition. Previously accumulated degradation is therefore mapped to an equivalent exposure under the current stress coefficient, $k_t^{cal} = k_T(T)k_{SOC}(SOC_t)$:

$$t_t^* = \left(\frac{Q_{t-1}^{cal}}{k_t^{cal}}\right)^{1/z} \tag{10}$$

The calendar loss is then updated as

$$Q_t^{cal} = k_t^{cal}(t_t^* + \Delta t_s)^z \tag{11}$$

where $\Delta t_s = 3{,}600\ s$ for the hourly chronology. The virtual exposure $t_t^*$ preserves accumulated degradation as SOC stress changes without altering physical elapsed time.

#### 3.3.2 Cycle Aging and Rainflow Counting

Cycle-induced capacity loss follows the complementary formulation in [5]:

$$Q^{cyc} = k_T^{cyc} k_C(C) k_{DoC}(DoC) FEC^z, \qquad z = 0.5 \tag{12}$$

where $C$ is cycle C-rate, $DoC$ is depth of cycle, and $FEC$ denotes cumulative full equivalent cycles. At the reference temperature, the temperature coefficient equals unity. The fitted C-rate and $DoC$ stress relationships are

$$k_C(C) = 0.0630C + 0.0971 \quad (13)$$

$$k_{DoC}(DoC) = 4.0253(DoC - 0.6)^3 + 1.0923 \quad (14)$$

Because optimized BESS operation contains irregular partial cycles, cycle depths are extracted from the executed SOC trajectory using rainflow counting according to ASTM E1049 [23]. Each detected cycle $j$ contributes $\Delta FEC_j = n_j DoC_j$, where $n_j = 1$ for a complete cycle and 0.5 for a residual half cycle. An effective C-rate is assigned to each detected cycle using only the periods in which the BESS is actively charging or discharging:

$$C_j^{eff} = \frac{1}{\left|\Omega_j^{act}\right|} \sum_{t \in \Omega_j^{act}} \frac{P_t^{ch} + P_t^{dch}}{E_{b(t)}^{avail}} \quad (15)$$

where $\Omega_j^{act}$ contains only the periods in which the BESS is actively charging or discharging, and $b(t)$ denotes the health-update interval containing period $t$. Excluding idle periods prevents rest hours from artificially reducing the effective cycle rate. Each active period is normalized by the available energy at that time, preserving the C-rate history of cycles spanning multiple health-update intervals.

Because the cycle stress coefficient varies among detected cycles, accumulated cycle loss is mapped to equivalent exposure under the current stress:

$$FEC_j^* = \left(\frac{Q_{j-1}^{cyc}}{k_j^{cyc}}\right)^{1/z} \quad (16)$$

and updated according to

$$Q_j^{cyc} = k_j^{cyc}\left(FEC_j^* + \Delta FEC_j\right)^z \quad (17)$$

Rainflow residuals are carried across health-update intervals and reset only at physical battery replacement. The square-root exposure relationship also means that equal FEC increments need not produce equal incremental capacity loss, distinguishing the reference model from the linear-throughput surrogate examined later. Calendar and cycle losses are added to determine battery SOH:

$$\begin{gathered} Q_b^{tot} = Q_b^{cal} + Q_b^{cyc} \\ SOH_b = 1 - \frac{Q_b^{tot}}{100} \end{gathered} \quad (18)$$

The available energy for the subsequent operating interval is then $E_b^{avail} = SOH_b S^B$, and replacement is triggered when

$$SOH_b \leq SOH^{EOL} \quad (19)$$

Replacement is evaluated at each health-update boundary. When the threshold is reached, SOH and available capacity return to BOL values, and calendar, cycle, and rainflow-residual histories are reset.

### *3.4 Operational Battery Performance and Linearization*

During lifecycle validation, charging/discharging efficiency and self-discharge are updated with SOH so that both available capacity and operating losses evolve with battery condition.

### *3.4.1 C-Rate- and SOH-Dependent Efficiency*

The fixed efficiencies used during planning are replaced during lifecycle validation by separate charging and discharging efficiency curves derived from [6]. Retaining the one-way curves preserves the source asymmetry rather than imposing a symmetric allocation of round-trip losses. For a battery with available energy $E_b^{avail}$, the instantaneous operating rate is

$$C_{b,t} = \frac{P_{b,t}^{ch} + P_{b,t}^{dch}}{E_b^{avail}} \tag{20}$$

Consequently, the same MW operating point corresponds to a higher C-rate as usable capacity declines. A system-level BOL RTE of 85%, adopted from the NREL Annual Technology Baseline [24], is imposed at the 1C normalization point. The source one-way efficiency curves are scaled using

$$s_\eta = \sqrt{\frac{\eta_{1C}^{RTE,sys}}{\eta_{1C}^{ch,cell} \eta_{1C}^{dch,cell}}} \tag{21}$$

Using $\eta_{1C}^{RTE,sys} = 0.85$, $\eta_{1C}^{ch,cell} = 0.961$, and $\eta_{1C}^{dch,cell} = 0.947$, gives $s_\eta = 0.96644$. The normalized BOL one-way efficiencies are

$$\begin{aligned} \eta_0^{ch}(C_i) &= s_\eta \eta^{ch,cell}(C_i) \\ \eta_0^{dch}(C_i) &= s_\eta \eta^{dch,cell}(C_i) \end{aligned} \tag{22}$$

This preserves the source curve shapes and charging/discharging asymmetry while setting their 1C product to the adopted system RTE. The normalized curves are shown in Fig. 3.

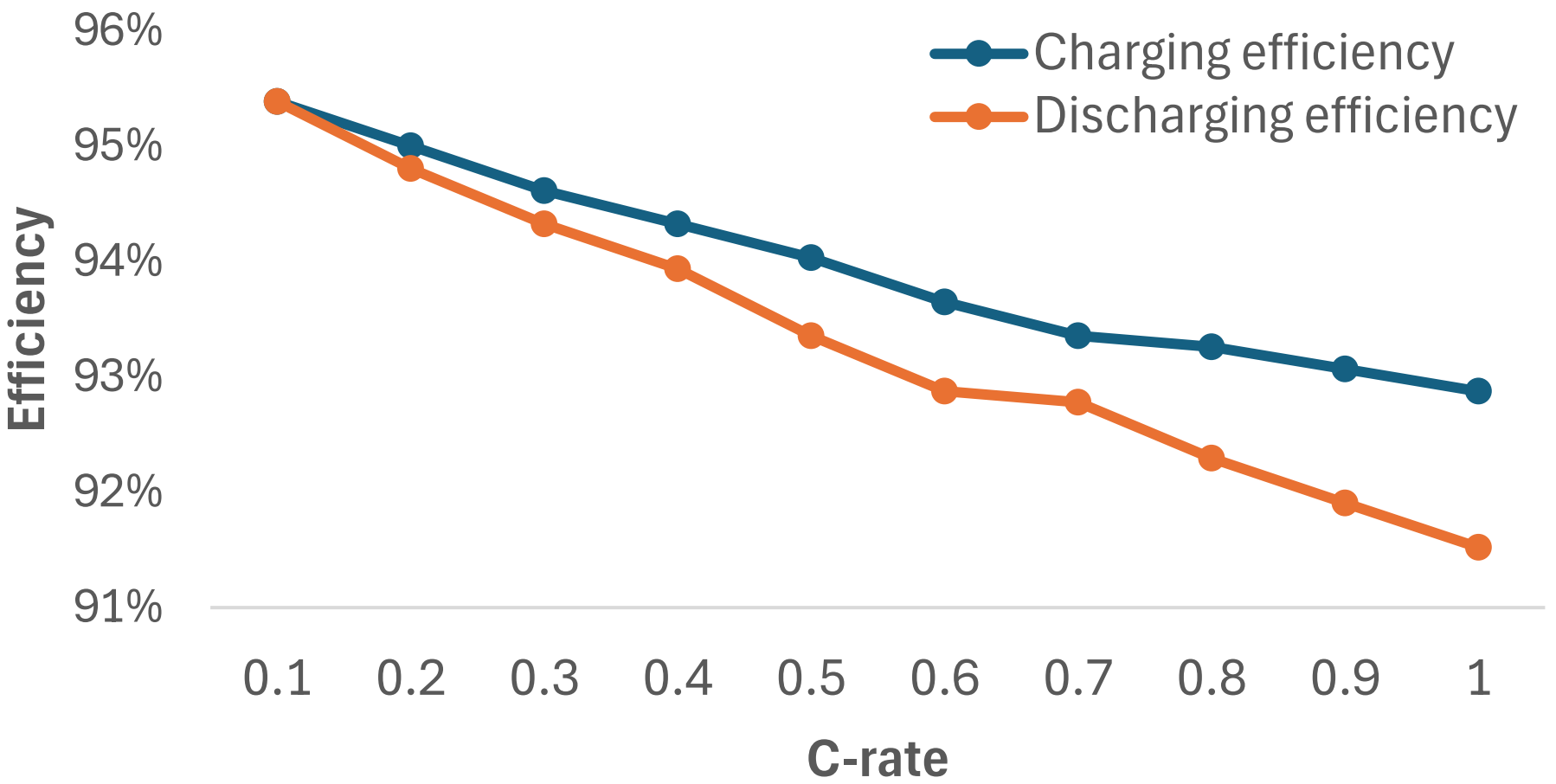

Fig. 3. Normalized BOL charging and discharging efficiencies versus C-rate based on [6]; their product equals the adopted 0.85 RTE at 1C.

Battery aging also modifies the efficiency level. Based on the normalized trend in [25], the SOH-dependent relationship is approximated as

$$\eta^{src}(SOH) = 0.95 + 0.03333(SOH - 1) \tag{23}$$

The corresponding normalized multiplier is

$$m_\eta(SOH) = \frac{\eta^{src}(SOH)}{\eta^{src}(1)} \tag{24}$$

The multiplier is applied while preserving the underlying C-rate dependence:

$$\begin{aligned} \eta^{ch}(C, SOH) &= \eta_0^{ch}(C)\sqrt{m_\eta(SOH)} \\ \eta^{dch}(C, SOH) &= \eta_0^{dch}(C)\sqrt{m_\eta(SOH)} \end{aligned} \tag{25}$$

Only the additional SOH effect is split symmetrically between charging and discharging; the underlying one-way asymmetry is retained. At the 80% SOH replacement threshold, the reference relationship reduces the 1C system RTE from 0.850 to approximately 0.844. Stronger SOH-dependent efficiency sensitivities are evaluated in Sections 4.2 and 5.2.

### *3.4.2 Self-Discharge*

A BOL self-discharge value of 3% per month is adopted from the manufacturer-specified upper limit for the LFP cell in [26] and converted to an equivalent hourly rate. Although self-discharge can vary with battery condition [20], a transferable stationary-LFP relationship could not be established. The reference case therefore adopts a study-specific sensitivity in which monthly self-discharge increases linearly from 3% at BOL to 6% at the 80% SOH replacement threshold:

$$s_b^m(SOH) = s_0^m\left[1 + \frac{1-SOH}{1-SOH^{EOL}}\right], \qquad s_0^m = 0.03. \tag{26}$$

This relationship is treated as an assumption rather than a generally validated LFP degradation law and is isolated through the ablation in Sections 4.2 and 5.2.

### *3.4.3 Piecewise Linear Implementation of Dynamic Efficiency*

Direct multiplication of optimized battery power by rate-dependent efficiency would introduce nonlinear terms. Because SOH and available energy are fixed within each health-update interval, the efficiency curves are converted to fixed power-to-energy-flow maps before solving that interval.

For each C-rate breakpoint $C_i$, the corresponding power breakpoint is

$$P_{b,i} = C_i E_b^{avail} \tag{27}$$

A zero-power breakpoint is included explicitly. The segment from 0–0.1C breakpoint uses the 0.1C charging and discharging efficiencies, yielding constant one-way efficiency over that range. Cases using C-rate-dependent efficiency are limited to the maximum adopted C-rate relative to current available energy to prevent extrapolation beyond the source curves; cases without it retain the fixed 1 h rated-power limit.

The effective stored-energy gain and withdrawal are

$$\begin{aligned} G_{b,i} &= P_{b,i}\eta^{ch}(C_i, SOH_b) \\ W_{b,i} &= \frac{P_{b,i}}{\eta^{dch}(C_i, SOH_b)} \end{aligned} \tag{28}$$

The optimization therefore interpolates the mappings $P^{ch} \rightarrow G$ and $P^{dch} \rightarrow W$ directly, rather than interpolating efficiency and multiplying it by power.

Charging is represented using special ordered set of type 2 (SOS2) interpolation:

$$\begin{aligned} P_{b,t}^{ch} &= \sum_i \lambda_{b,t,i}^{ch} P_{b,i} \\ G_{b,t} &= \sum_i \lambda_{b,t,i}^{ch} G_{b,i} \\ \sum_i \lambda_{b,t,i}^{ch} = 1, &\qquad \lambda_{b,t,i}^{ch} \geq 0, \end{aligned} \tag{29}$$

with at most two adjacent weights active. The corresponding discharge mapping is

$$\begin{aligned} P_{b,t}^{dch} &= \sum_i \lambda_{b,t,i}^{dch} P_{b,i} \\ W_{b,t} &= \sum_i \lambda_{b,t,i}^{dch} W_{b,i} \\ \sum_i \lambda_{b,t,i}^{dch} = 1, &\qquad \lambda_{b,t,i}^{dch} \geq 0, \end{aligned} \tag{30}$$

Thus, the stored energy equation can then be written as

$$E_{b,t+1}^{B} = (1 - \sigma_b) E_{b,t}^{B} + G_{b,t} \Delta t - W_{b,t} \Delta t, \qquad 0 \leq E_{b,t}^{B} \leq E_b^{avail} \tag{31}$$

The resulting formulation preserves continuous operation between neighboring efficiency breakpoints while remaining compatible with mixed-integer linear optimization.

## *3.5 Sequential Health Update and Lifecycle Validation*

Lifecycle validation proceeds sequentially over battery health-update intervals. At each boundary, the inherited physical state is fixed and the operating problem is solved from that point to the end of the current lifecycle year; only the dispatch within the current health interval is executed. Degradation and performance are then updated before the remaining year is reoptimized.

The full-chronology simulation starts at 50% SOC. No terminal SOC constraint is imposed at health-update boundaries; each rolling solve retains the 50% year-end condition. During ordinary updates, the SOC fraction is preserved while stored energy is rescaled to the updated available capacity. At replacement, the pre-replacement SOC fraction is transferred to the new battery while SOH, degradation exposure, and rainflow histories are reset.

Calendar and cycle degradation are evaluated using (7)–(17), SOH is updated through (18), and available capacity, efficiency, self-discharge, and the power breakpoints in (27) are reconstructed for the next interval. If the criterion in (19) is reached, replacement occurs; otherwise the updated state is carried forward through the 20-year horizon.

Lifecycle economic performance is evaluated using net present cost (NPC). Because the installed portfolio is fixed, differences among fidelity cases arise from operating costs and battery replacement rather than capacity resizing. Let $C_y^{op}$ denote the annual operating cost comprising net grid transactions and PV fixed O&M, and $C_k^{rep}$ the replacement cost of event $k$ occurring at time $\tau_k$, measured in years from project start.

$$NPC = c^{PV} S^{PV} + c^B S^B + \sum_{y \in Y} \frac{C_y^{op}}{(1+r)^{y-1}} + \sum_{k \in K^{rep}} \frac{C_k^{rep}}{(1+r)^{\tau_k}}. \tag{32}$$

The reliability-enforcement term in (1) is excluded from the reported economic NPC. Energy adequacy is instead evaluated directly using cumulative energy not served,

$$ENS = \sum_{b \in B} \sum_{t \in T_b} P_{b,t}^{shed} \Delta t \tag{33}$$

and loss of load hours (LOLH),

$$LOLH = \sum_{b \in B} \sum_{t \in T_b} \mathbb{I}\left(P_{b,t}^{shed} > \varepsilon\right) \Delta t \tag{34}$$

where $\mathbb{I}(\cdot)$ is an indicator function and $\epsilon = 10^{-6}$ MW is the numerical threshold used to identify a nonzero shortfall; $\mathbb{I}\left(P_{b,t}^{shed} > \epsilon\right) = 1$ when the condition is satisfied and 0 otherwise. Shortfall days, $N^{SF}$, count calendar days containing at least one period with unserved load exceeding $\epsilon$, while peak shortfall is defined as $P_{\max}^{SF} = \max_{b,t} P_{b,t}^{shed}$. These complementary measures distinguish the total magnitude of unmet energy from the duration, frequency, and severity of individual adequacy events.

# 4. Case Description and Experimental Design

## *4.1 System Data and Reference Assumptions*

The fidelity assessment uses a grid-connected commercial microgrid in Houston, Texas, with utility exchange, PV generation, and an LFP/graphite BESS connected at a common node. The single-node representation isolates battery and temporal modeling effects from distribution-network constraints. A 20-year horizon and 3% real discount rate are considered.

Year-1 demand uses a Houston commercial-building profile scaled to a 10 MW peak [27], and grows by 0.5% annually. The microgrid has a 5 MW point of common coupling (PCC). PV availability is obtained from an hourly Houston NREL PVWatts profile [28] and degrades by 0.5% annually. Grid import prices follow the 2022 Houston locational marginal price chronology from the U.S. Energy Information Administration ERCOT dataset [29], while exports are valued at 80% of the corresponding import price. The PV and BESS capital costs are $c^{PV} = \$1.12$ million/MW and $c^{B} = \$0.684$ million/MWh, respectively; annual fixed operation and maintenance (O&M) is \$19,000/MW-year for PV and set to zero for BESS. Each BESS replacement is assumed to cost 80% of the initial BESS capital cost and is discounted at its occurrence time. No terminal residual or salvage value is credited. These economic assumptions are common to all fidelity cases.

The planning model in Section 3.2 yields $S^{PV} = 29.515$ MW and $S^{B} = 52.196$ MWh. With the fixed 1 h architecture, rated BESS power is 52.196 MW. Battery investment is modeled on an energy-capacity basis, so power capacity is not independently optimized or costed. The resulting capacities remain fixed throughout lifecycle validation.

The reference BESS begins at 100% SOH with an 85% system-level BOL RTE at 1C. Battery temperature is fixed at $25°\mathrm{C}$, BOL self-discharge is 3% per month, and replacement occurs at 80% SOH. The reference lifecycle case uses all 8,760 hourly periods and a three-month health-update interval; SOH, available energy, efficiency maps, and self-discharge remain fixed within each interval.

One 8,760-h load, PV, and price chronology is repeated over the 20-year horizon with deterministic load growth and PV degradation. Thus, adequacy-critical calendar days identified later are specific to the adopted chronology; the temporal analysis concerns preservation of adequacy-relevant operating conditions rather than particular dates.

## *4.2 Battery Fidelity, Ablation, and Sensitivity Cases*

The battery hierarchy progressively activates lifecycle mechanisms while retaining the same installed PV and BESS capacities, as summarized in Table 2.

Table 2 Battery fidelity hierarchy

| *Case* | *Capacity degradation* | *Cycle-specific C-rate* | *C-rate-dependent efficiency* | *SOH-dependent efficiency* | *SOH-dependent self-discharge* |
|---|---|---|---|---|---|
| B0 | No | No | No | No | No |
| B1 | Calendar + cycle | No | No | No | No |
| B2 | Calendar + cycle | Yes | Yes | No | No |
| B3 | Calendar + cycle | Yes | Yes | Yes | Yes |

B1 activates nonlinear calendar and rainflow-based cycle aging, capacity loss, and replacement while using a fixed cycle-aging C-rate of 0.09399C and the same fixed 0.85 system RTE used in planning. B2 adds cycle-specific C-rate and C-rate-dependent charging/discharging efficiency, while B3 further includes SOH-dependent efficiency and self-discharge and serves as the reference representation. The 0.09399C fixed rate is calculated from active B0 operation and differs from the 0.0837C first-life rainflow-effective rate used to calibrate the cycle-only linear surrogate because the two use different averaging procedures.

Two linear degradation surrogates are evaluated separately. The cycle-only throughput surrogate is calibrated from B3 first-life operation using a representative rainflow depth of cycle of 0.5954 and effective C-rate of 0.0837C, giving a Naumann cycle-aging coefficient of 0.111823 and approximately 31,989 FEC for 20% cycle-only capacity loss. Because it omits calendar aging while also linearizing cycle degradation, this surrogate confounds mechanism omission with linearization. A second surrogate therefore retains calendar and cycling degradation while linearizing their accumulation:

$$Q_{\text{loss}}^{\text{lin}} = Q_{\text{cal}}^{\text{lin}} + Q_{\text{cyc}}^{\text{lin}} = \alpha_{\text{cal}} T_{\text{age}} + \alpha_{\text{cyc}} FEC^{\text{throughput}}, \quad FEC^{\text{throughput}} = \frac{1}{2S^B} \sum_t \left(P_t^{ch} + P_t^{dch}\right) \Delta t \tag{35}$$

where $T_{\text{age}}$ is the elapsed time since battery commissioning and $FEC^{\text{throughput}}$ is the cumulative energy throughput expressed as equivalent full cycles using the original installed BESS energy capacity; both reset at a battery replacement. The calendar and cycling coefficients are calibrated separately to the corresponding B3 losses at the first replacement boundary. At year 9, B3 records 12.6068% calendar loss and 7.6602% cycle loss after 78,840 h and 3059.13 throughput FEC. This gives $\alpha_{\text{cal}} = 1.5990 \times 10^{-4}\%/\text{h}$, equivalent to 1.40075 percentage points/year, and $\alpha_{\text{cyc}} = 2.50405 \times 10^{-3}\%/\text{FEC}$. These coefficients are then fixed for the remaining lifecycle, while all other B3 settings are retained.

The progressive hierarchy measures the cumulative effect of increasing battery-model fidelity, whereas targeted ablations isolate individual mechanisms around B3 by removing calendar aging, replacing cycle-specific C-rate with the fixed application-average rate, or removing C-rate-dependent efficiency, SOH-dependent efficiency, or SOH-dependent self-discharge. All other B3 settings remain unchanged.

To separate efficiency-level effects from the remaining operating-efficiency representation, two C-rate-independent controls are included. The application-weighted BOL RTE under B3 is 0.90397; the controls therefore use BOL RTE values of 0.90397 and 0.85, split symmetrically between charging and discharging while retaining the B3 SOH-dependent multiplier. The matched 0.90397 control isolates the residual effect of replacing the C-rate-dependent asymmetric one-way maps with a C-rate-independent symmetric representation, whereas the 0.85 control additionally includes the lower average efficiency level.

B3 uses the mild SOH-dependent efficiency relationship in Section 3.4.1. Moderate and high sensitivity cases retain a 1C BOL RTE of 0.85 but reduce RTE at the 80% SOH threshold to 0.830 and 0.800, respectively, compared with 0.844 under B3. Their efficiency multipliers vary linearly with SOH while preserving the C-rate-dependent curve shape. These are prescribed sensitivity cases rather than experimentally established stationary-LFP aging trajectories.

### *4.3 Health Update and Temporal Representation Tests*

The reference B3 model updates battery health and reoptimizes operation every three months. Otherwise identical cases use 6- and 12-month intervals; the three-month case is a reference rather than a formally converged minimum. Longer intervals hold SOH, available energy, efficiency, and self-discharge fixed for longer while reducing redispatch frequency, so the experiment captures their combined effect. Replacement is evaluated at the end of each health interval.

Temporal fidelity is evaluated using the full 8,760-h chronology, monthly- and annual-average profiles, and three 12-day clustered representations. Monthly averaging produces one synthetic 24 h profile per month by hourly averaging load, PV availability, and import price, with each profile weighted by the number of days in that month. Annual averaging similarly produces one 24 h profile weighted by 365. All representative-day cases impose 50% SOC at the beginning and end of each modeled day.

For clustered cases, the 24 h load, PV-availability, and import-price profiles are standardized by feature and concatenated into 72-element vectors. Euclidean-distance k-medoids selects 11 actual days, allowing one explicit extreme day to be added within a common 12-day computational budget. The energy-deficit metric used to identify the alternative extreme day is

$$D_d = \sum_{t \in \mathcal{H}_d} \max\left(P_{1,t}^{L} - S^{PV}\phi_t^{PV} - P^{PCC}, 0\right) \Delta t \tag{36}$$

where $\mathcal{H}_d$ contains the hourly periods in day $d$. Thus, $D_d$ measures the year-1 net-load energy above the PCC import limit before BESS operation and is used only as a temporal-stress metric.

In the peak-informed cases, the maximum instantaneous net-load day is added as the twelfth day. Under natural weighting, each medoid receives the number of source days assigned to its cluster using quarter-specific exposure counts, while the preserved peak day has unit weight. The calibrated case instead adjusts integer weights to minimize normalized mismatch in annual load energy, PV availability, and mean import price while preserving 365 total days, exact quarterly counts, all selected days, and unit weight for the preserved extreme. The energy-deficit-informed case uses the same 11 medoids but replaces the peak-net-load day with the maximum-$D_d$ day and recalibrates the weights identically.

Representative-day operation and degradation are weighted by the corresponding day counts, but independent daily SOC closure removes original inter-day state continuity. The reduced cases therefore test complete temporal representations rather than clustering alone. To diagnose this effect, the full B3 trajectory is additionally examined across Days 351–353 in lifecycle years 1 and 9.

### 4.4 Combined Simplification, Implementation, and Verification

To examine interactions among individually modest approximations, a combined case retains the full 8,760-h chronology while increasing the health-update interval from 3 to 6 months, replacing cycle-specific C-rate with the B0 application-average rate of 0.09399C, and disabling SOH-dependent efficiency and self-discharge. C-rate-dependent charging/discharging efficiency, calendar aging, rainflow-resolved cycling, cycle-depth dependence, BOL self-discharge, and the 80% replacement threshold are retained. The case is intended to test approximation interactions rather than define a universal minimum-fidelity model.

The complete experimental design is summarized in Table 3. Models are implemented in Python/Pyomo and solved with Gurobi 12.0.3 on a 12th Gen Intel Core i5-12450H processor with 8 GB RAM. Planning uses a 0.5% relative MIP-gap target and 7,200-s limit; lifecycle solves use the same gap target and a 3,600-s limit. The B3 reference case was additionally re-solved under alternative solver tolerances and settings. Across four robustness runs, cumulative ENS varied by only 0.0101 MWh, while replacement timing and the principal event-based adequacy metrics were unchanged. Because lifecycle solves use a 0.5% relative MIP-gap target, smaller NPC deviations are reported for completeness but are treated as numerically indistinguishable from the reference rather than interpreted directionally.

Lifecycle checks verify hourly power balance and BESS energy evolution, reconcile calendar and cycle losses with SOH, confirm replacement events against the 80% threshold, reconstruct ENS from hourly unserved load, and verify representative-day annual and quarterly weights. The load-shedding enforcement coefficient is excluded from economic NPC; adequacy is evaluated separately using ENS, LOLH, shortfall days, and peak shortfall. The full 8,760-h B3 case is the reference, so reported deviations measure differences from this modeling representation rather than physical ground truth.

Table 3 Fidelity experiments

| ***Experiment*** | ***Cases*** | ***Purpose*** |
|---|---|---|
| Battery hierarchy | B0, B1, B2, B3 | Cumulative battery fidelity |
| Linear surrogate | Cycle-only throughput; calibrated calendar + throughput | Separate omission from linearization |
| Targeted ablations | Calendar aging; cycle-specific C-rate; C-rate-dependent efficiency; SOH-dependent efficiency; SOH-dependent self-discharge | Isolate individual mechanisms |
| Efficiency controls | C-rate-dependent; C-rate-independent BOL RTE = 0.90397 and 0.85 | Separate efficiency-level and residual representation effects |
| SOH-dependent efficiency sensitivity | Mild, moderate, high | Test stronger deterioration |
| Health update interval | 3, 6, 12 months | Test health feedback and redispatch frequency |
| Temporal representation | 8,760 h; monthly/annual average; natural-weight cluster; peak- and energy-deficit-informed calibrated clusters | Test reduction, extremes, and state continuity |
| Combined simplification | Selected battery simplifications + 6-month update | Examine interaction among modest simplifications |

## 5. Results and Discussion

The results are evaluated against the full 8,760-h B3 lifecycle representation. Comparisons consider lifecycle economics, battery degradation and replacement, energy adequacy, and computational burden. Because B3 is a modeling reference rather than physical ground truth, differences relative to B3 are interpreted as deviations from the adopted reference representation.

### *5.1 Reference Lifecycle Behavior and Battery Fidelity Hierarchy*

Under B3, the fixed portfolio reaches the 80% SOH replacement threshold in years 9 and 18 and ends the 20-year horizon at 90.59% SOH. Lifecycle NPC increases from the degradation-naive planning estimate of $72.763 million to $111.255 million, including approximately $38.667 million in battery replacement cost.

B3 produces 24.012 MWh of cumulative ENS, equivalent to 0.002588% of lifecycle demand. Shortfalls occur over 29 loss-of-load hours on eight calendar days, with a maximum hourly deficit of 2.261 MW. ENS is therefore used primarily as a stress-sensitive measure of whether alternative fidelity representations preserve adequacy-critical operating conditions rather than as evidence of severe inadequacy in the fixed portfolio.

Table 4 compares the battery hierarchy and linear degradation surrogates.

Table 4 Lifecycle outcomes across battery fidelity cases

| *Case* | *Replacement time (yr)* | *ENS (MWh)* | *NPC deviation vs. B3* | *Final SOH* |
|---|---|---|---|---|
| B0 | None | 0.00 | -34.60% | 1.000 |
| B1 | 9.25 / 18.50 | 37.93 | +0.40% | 0.920 |
| B2 | 9.00 / 18.00 | 23.07 | -0.06% | 0.906 |
| B3 | 9.00 / 18.00 | 24.01 | Reference | 0.906 |
| Cycle-only linear throughput | None | 0.56 | -35.39% | 0.956 |
| Calibrated linear calendar + throughput | 9.00 / 18.00 | 16.38 | -0.20% | 0.955 |

B1 reproduces the need for replacement but produces substantially more ENS than B3 despite an NPC deviation of only 0.40%. B2 closely reproduces replacement timing and lifecycle cost, with ENS also close to B3. Thus, convergence in lifecycle NPC does not necessarily imply equivalent adequacy behavior.

The linear surrogates distinguish degradation-mechanism omission from linearization. The cycle-only surrogate predicts no replacement and an NPC 35.39% below B3. In contrast, the calibrated calendar-plus-throughput surrogate reproduces replacements in years 9 and 18 and limits the NPC deviation to −0.20%, although its 16.38 MWh ENS remains 31.8% below B3.

This contrast shows that the large lifecycle-cost error of the cycle-only surrogate is driven primarily by omission of calendar aging rather than linearization alone. At the first B3 replacement boundary, calendar aging contributes 12.61 percentage points, or 62.2% of accumulated capacity loss, compared with 7.66 percentage points from cycling. Retaining both mechanisms therefore enables a first-life endpoint-calibrated linear model to reproduce replacement timing and lifecycle cost closely, although the complete SOH trajectory and adequacy outcome are not preserved.

### *5.2 Mechanism-Specific Effects and Adequacy-Metric Dependence*

Targeted ablations isolate individual mechanisms around B3. Calendar aging has the largest effect on replacement and lifecycle economics in this application: removing it eliminates both replacements, decreases NPC from $111.255 million to $72.181 million, and shifts the first year with ENS from 9 to 16. Cycle aging alone leaves the battery at approximately 88.96% SOH after 20 years, above the 80% replacement threshold.

Fig. 4 further shows that B3 reaches the replacement threshold in years 9 and 18, whereas nonlinear cycle aging alone and the cycle-only linear surrogate remain above the threshold throughout the horizon. The calibrated calendar-plus-throughput surrogate reproduces both replacement times but follows a different within-life trajectory, ending at approximately 95.5% SOH versus 90.6% under B3. Thus, retaining the dominant degradation mechanisms can preserve replacement timing without preserving the complete SOH trajectory. This result is specific to the operating duty examined here.

The remaining battery refinements have smaller individual effects around B3. Replacing cycle-specific C-rate with the fixed application-average rate decreases ENS by 0.156 MWh, while removing SOH-dependent efficiency and self-discharge decreases ENS by 0.848 and 0.097 MWh, respectively. These modest effects are consistent with the low operating duty: the mean active C-rate under B3 is approximately 0.098C, far below the 1C rated capability of the 52.196 MWh/52.196 MW architecture. The limited cycle-C-rate effect should therefore be interpreted as duty-specific rather than generally negligible.

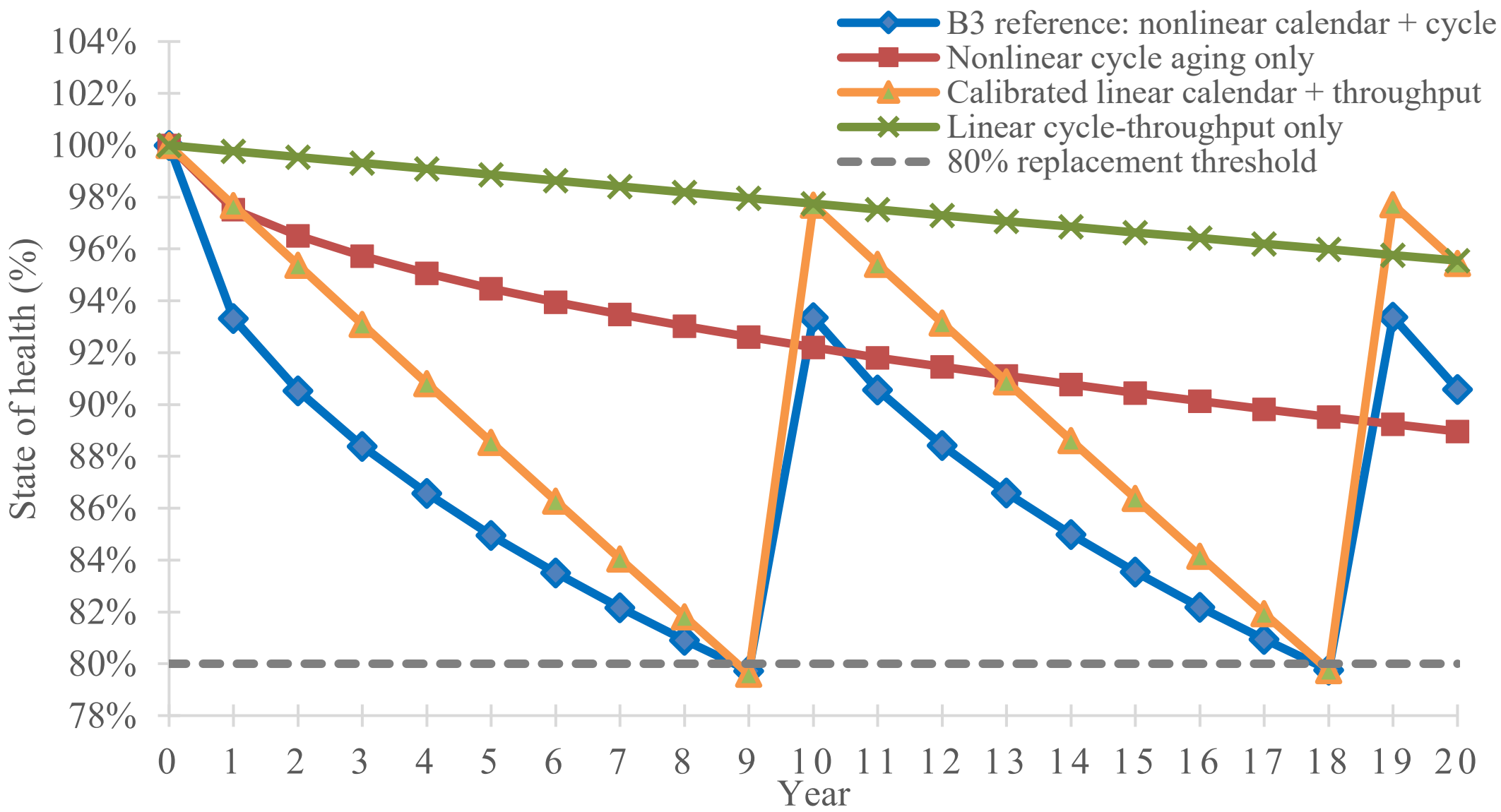

Fig. 4. BESS SOH trajectories under alternative degradation representations; dashed line: 80% replacement threshold.

Replacing the C-rate-dependent asymmetric efficiency maps with a C-rate-independent BOL RTE changes both efficiency level and operating-rate dependence. The application-weighted BOL RTE under B3 is 0.90397. Relative to the 0.85 control, the matched 0.90397 control indicates that approximately 13.607 MWh, or 92.3%, of the ENS difference is associated with the efficiency-level change, while the remaining 1.142 MWh, or 7.7%, reflects the residual operating-efficiency representation, including loss of C-rate dependence and one-way asymmetry. These proportions apply only to this controlled comparison. The adequacy metrics respond differently, as summarized in Table 5.

Table 5 Adequacy outcomes under the efficiency controls

| *Case* | *ENS (MWh)* | *LOLH (h)* | *Shortfall days* | *Peak shortfall (MW)* |
|---|---|---|---|---|
| B3 reference | 24.01 | 29 | 8 | 2.26 |
| C-rate independent, BOL RTE = 0.90397 | 25.15 | 18 | 10 | 2.70 |
| C-rate independent, BOL RTE = 0.85 | 38.77 | 24 | 14 | 3.22 |

The matched control produces slightly more ENS than B3 but fewer LOLH, while shortfall days and peak magnitude increase. The 0.85 control similarly produces greater ENS but fewer LOLH than B3. Thus, fewer affected hours can coincide with more frequent or severe deficits, and approximation quality depends on the adequacy metric examined.

Stronger SOH-dependent efficiency deterioration is also tested using the system RTE at 1C and 80% SOH. Table 6 shows that replacement timing is unchanged while adequacy becomes increasingly sensitive to stronger deterioration.

Table 6 SOH-dependent efficiency sensitivity

| *Case* | *RTE at 80% SOH (1C)* | *Replacement time (yr)* | *ENS (MWh)* | *LOLH / shortfall days* | *NPC deviation vs. B3* |
|---|---|---|---|---|---|
| Mild reference | 0.844 | 9 / 18 | 24.01 | 29 / 8 | Reference |
| Moderate | 0.830 | 9 / 18 | 26.46 | 31 / 10 | +0.141% |
| High | 0.800 | 9 / 18 | 32.63 | 38 / 12 | +0.447% |

The high case increases cumulative ENS by approximately 36% while changing lifecycle NPC by less than 0.5%. Replacement timing remains unchanged because the criterion depends on SOH, whereas the sensitivity changes conversion efficiency without directly altering the degradation trajectory. The moderate and high cases are prescribed sensitivities rather than experimentally established stationary-LFP aging trajectories.

### *5.3 Effect of the Health Update Interval*

Increasing the health update interval reduces the number of sequential operating problems but also changes how frequently battery condition is fed back into the model and dispatch is reoptimized. Table 7 compares the three tested intervals under otherwise identical B3 settings and full hourly chronology. All three intervals reproduce the same replacement times, while lifecycle NPC remains numerically indistinguishable from B3 within solver tolerance. Adequacy behavior is less stable. Under B3, the first shortfall occurs in year 9 and contributes 0.234 MWh of ENS. The 12-month case does not detect this early event and shifts the first observed shortfall to year 14, while cumulative ENS decreases by approximately 18.8%.

These lower ENS values do not represent an improvement in the physical portfolio, which remains unchanged. A longer interval simultaneously holds the battery health state fixed for longer and reduces redispatch frequency; the experiment therefore measures their combined effect. The results show that similar lifecycle cost and replacement timing across update intervals do not guarantee similar estimates of when adequacy problems emerge. The computational tradeoff is substantial. Moving from three- to six-month updates reduces runtime by 43.5%, while annual updating reduces it by 58.0%. The appropriate interval therefore depends on the lifecycle outcome that must be preserved rather than on agreement in lifecycle cost alone.

Table 7 Effect of the battery health update interval

| *Interval* | *ENS (MWh)* | *Replacement time (yr)* | *First ENS year* | *NPC deviation vs. B3* | *Runtime change vs. B3* |
|---|---|---|---|---|---|
| 3 months | 24.012 | 9 / 18 | 9 | Reference | Reference |
| 6 months | 22.182 | 9 / 18 | 9 | −0.020% | −43.5% |
| 12 months | 19.504 | 9 / 18 | 14 | −0.079% | −58.0% |

## *5.4 Temporal Representation and Adequacy*

Temporal reduction provides the largest computational savings and the strongest sensitivity in adequacy outcomes. Monthly-average, annual-average, natural-weight, and peak-informed calibrated representations all report zero ENS, compared with 24.012 MWh under the full 8,760-h B3 chronology. The energy-deficit-informed case instead produces 365.99 MWh despite using the same 12-day computational budget. Table 8 summarizes the lifecycle results.

Table 8 Lifecycle outcomes under temporal reduction

| *Representation* | *Replacement time (yr)* | *ENS (MWh)* | *NPC deviation vs. B3* | *Runtime change vs. B3* |
|---|---|---|---|---|
| Full 8,760 h chronology | 9.00 / 18.00 | 24.01 | Reference | Reference |
| 12-day cluster, peak-calibrated | 9.00 / 18.00 | 0.00 | +2.64% | −92.8% |
| 12-day cluster, deficit-calibrated | 9.00 / 18.00 | 365.99 | +2.59% | −93.0% |
| 12-day cluster, peak-natural | 8.75 / 17.50 | 0.00 | +1.03% | −91.8% |
| Monthly average | 8.50 / 17.00 | 0.00 | +1.58% | −96.2% |
| Annual average | 8.50 / 17.25 | 0.00 | +3.74% | −98.7% |

The two calibrated clustered cases isolate extreme-day selection: they use the same 11 k-medoids, calibrated weighting, and battery representation, differing only in the preserved extreme day. Both reproduce B3 replacements in years 9 and 18 and yield nearly identical NPC deviations and runtime reductions, yet their adequacy outcomes diverge completely: the peak-informed case reports zero ENS, while the energy-deficit-informed case reports 365.99 MWh. Thus, similar economic, degradation, and computational performance does not imply similar adequacy behavior.

Fig. 5 helps explain this sensitivity. Day 352 has the largest year-1 energy deficit above the 5 MW PCC limit, approximately 34.53 MWh, but ranks only seventeenth by instantaneous peak net load and is therefore absent from the peak-informed set. Day 243 contains the maximum instantaneous net load but ranks only thirteenth by daily energy deficit, at approximately 16.52 MWh. All B3 shortfalls occur on repetitions of Day 352 during affected lifecycle years. For an energy-limited resource, preserving the maximum instantaneous power requirement therefore does not guarantee preservation of the most demanding energy event.

Selecting the stress day alone is also insufficient because the full chronology encounters Day 352 under different battery-state conditions from the representative-day formulation. In year 1, B3 enters Day 352 at 90.0% SOC, reaches 100%, and finishes at 17.6% without ENS. By year 9, available BESS energy has declined from 49.18 to 41.77 MWh; Day 352 begins at 87.2% SOC, reaches zero during the event, and produces the first 0.234 MWh shortfall over three hours before ending at 13.9% SOC. The preceding chronology therefore provides substantial battery-state preconditioning before the critical event.

The representative-day formulation instead fixes SOC at 50% at both the beginning and end of each selected day. The energy-deficit-informed case therefore removes the high incoming SOC available under full chronology while requiring the battery to restore its state by day end, causing Day 352 to substantially overstate cumulative ENS relative to B3. A targeted fourth-quarter replay in lifecycle years 1 and 9, initialized from the corresponding B3 end-of-Q3 state and solved with the same formulation and year-end SOC condition, reproduces the reference Q4 ENS and confirms the diagnosed Day 352 trajectories.

These results show that extreme-day selection and inter-day battery-state continuity must be considered jointly when reduced chronology is used for storage adequacy. Peak-informed selection can omit sustained energy-stress events, while isolated representation of an energy-critical day under daily SOC closure can exaggerate its severity.

Because one annual chronology is repeated over the study horizon, the temporal experiments do not capture inter-annual variability in load, solar availability, or electricity prices. The general finding is therefore the dependence of storage adequacy on sustained energy stress and chronological state continuity rather than the significance of Day 352 itself.

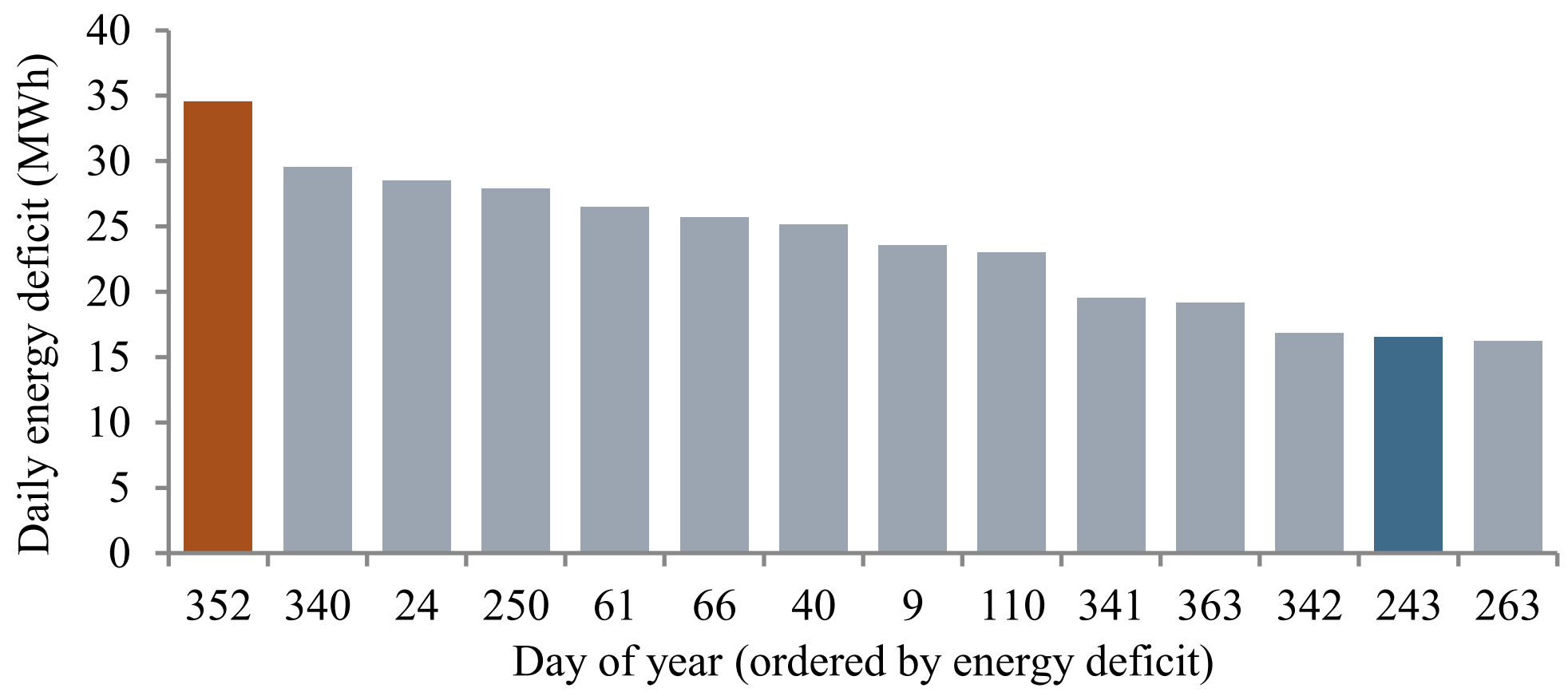

Fig. 5. Largest year-1 daily energy deficits above the PCC limit; Day 352 has the maximum energy deficit and Day 243 the maximum instantaneous net load.

### *5.5 Combined Simplification and Practical Fidelity Implications*

The combined simplification retains full hourly chronology and the principal calendar- and cycle-aging mechanisms while applying the selected battery and health-update approximations. Replacement remains in years 9 and 18, lifecycle NPC remains numerically indistinguishable from B3 within solver tolerance, and runtime decreases by approximately 40%. Adequacy is more sensitive: ENS decreases from 24.01 to 21.27 MWh, LOLH from 29 to 21 h, shortfall days from eight to six, and the first ENS year shifts from 9 to 14. The six-month-update-only case still detects ENS in year 9, indicating that individually modest approximations can interact to alter adequacy timing while leaving replacement and lifecycle economics nearly unchanged.

Across the experiments, fidelity requirements are metric-dependent. Retaining the dominant degradation mechanisms preserves replacement timing and lifecycle economics closely in this case, whereas SOH trajectory and energy adequacy remain more sensitive to representation. Temporal fidelity is particularly consequential: reduced cases with similar replacement timing, final SOH, NPC, and runtime can produce fundamentally different adequacy outcomes when extreme-event selection and inter-day battery-state continuity differ.

The results reflect the specific system configuration and operating assumptions considered in this study, including the 5 MW PCC, 1 h BESS architecture, repeated Houston chronology, fixed battery temperature, and deterministic inputs. PCC capacity and BESS duration are held fixed during lifecycle validation to isolate modeling-fidelity effects. The economic results also follow the adopted zero-salvage assumption, which slightly increases the contribution of the year-18 replacement to lifecycle NPC. Accordingly, the relative sensitivity to individual simplifications may vary under different operating conditions or system designs. Table 9 summarizes the principal implications.

Table 9 Practical modeling-fidelity implications

| ***Lifecycle outcome*** | ***Most consequential fidelity elements*** | ***Practical implication*** |
|---|---|---|
| Lifecycle cost | Calendar aging; replacement; retention of calendar and cycle degradation | Linear degradation can reproduce lifecycle cost closely when both aging mechanisms are retained and separately calibrated |
| Replacement timing | Calendar and cycle aging; replacement logic | The calibrated linear surrogate reproduces years 9 and 18; 3–12-month health updates do not change replacement years |
| SOH trajectory | Degradation functional form; calibration | Endpoint calibration can preserve replacement timing without preserving the within-life SOH trajectory |
| Energy adequacy | Temporal representation; inter-day SOC continuity; health-update frequency; operating losses | Extreme-event selection and state continuity must be considered jointly; agreement in NPC or replacement timing does not ensure adequacy fidelity |

## 6. Conclusion

This study assessed modeling fidelity for long-term BESS lifecycle evaluation using a common plan-freeze-validate-quantify framework. A fixed PV-BESS portfolio was evaluated under alternative degradation, operational, health-update, and temporal representations so that differences in lifecycle outcomes reflect modeling choices rather than changes in system design. The B3 reference produces replacements in years 9 and 18, a lifecycle NPC of \$111.255 million, and 24.012 MWh of cumulative ENS. Removing calendar aging eliminates both replacements and reduces NPC by approximately 35.1%. In contrast, a linear surrogate retaining separately calibrated calendar and throughput degradation reproduces both replacement years and limits the NPC deviation to 0.20%, although ENS remains 31.8% below B3 and the nonlinear SOH trajectory is not reproduced. Thus, retaining and calibrating the dominant degradation mechanisms can be more consequential for lifecycle economics than preserving nonlinear form alone. Health-update and temporal representations have stronger effects on adequacy. I Increasing the health-update interval from three to twelve months preserves replacement timing, with lifecycle NPC remaining numerically indistinguishable from B3 within solver tolerance, but shifts the first detected shortfall from year 9 to year 14. Under temporal reduction, the peak-informed calibrated 12-day case preserves replacement timing and reduces runtime by 92.8% while reporting zero ENS. Replacing the preserved peak-net-load day with the maximum daily-energy-deficit day leaves replacement timing, final

SOH, NPC, and runtime nearly unchanged but increases ENS to 365.99 MWh. Full-chronology diagnostics show that this divergence arises from differences in the battery state surrounding the critical event.

Overall, no single fidelity level preserves every lifecycle metric. Simplified degradation models can reproduce replacement timing and lifecycle economics when dominant aging mechanisms are retained and calibrated, but trajectory-sensitive outcomes require separate validation. Likewise, temporal reduction should be evaluated against both adequacy-critical events and the battery-state continuity surrounding them. Modeling fidelity should therefore be selected according to the lifecycle outcome the analysis is intended to preserve.